# A Joint Wireless-Optical Front-haul Solution for Multi-user Massive MIMO 5G RAN

Son T. Le[*,1], Stefan Wesemann[2], Roman Dischler[2] and Sivarama Venkatesan[1]

[1]Nokia Bell Labs, Holmdel, NJ, USA, son.thai_le@nokia-bell-labs.com
[2]Nokia Bell Labs, Stuttgart, Germany

**Abstract** *We demonstrate a high capacity IF-over-fiber mobile fronthaul solution for multi-user massive MIMO 5G RAN. Using this scheme, a record aggregated radio bandwidth of 25.6 GHz was transmitted on a single optical wavelength over 40 km without fiber chromatic dispersion compensation.*

## Introduction

Centralized radio access networks (C-RANs) have been widely deployed in LTE networks worldwide, greatly benefiting wireless network capacity and resulting OPEX savings compared to traditional D-RAN architectures [1]. Deploying C-RAN for 5G networks with massive MIMO using the conventional Common Public Radio Interface (CPRI) protocol is very challenging due to the prohibitively high capacity requirements for optical mobile front-haul (FH) links. To address this problem, a new specification called evolved (or enhanced) CPRI (eCPRI) has been introduced [2], including flexible functional splits at the physical layer between the remote radio unit (RRU) and the central unit (CU) [2]. Out of 8 function split options specified in eCPRI, the option 7 (intra PHY layer split) reduces the FH capacity by order of magnitude by moving the MIMO beamforming and a significant portion of the signal processing back to the RRU. This, however, poses several constraints to the wireless MIMO beamforming capability (e. g., a reduced number of spatial layers) and thus limits the cell's throughput and performance.

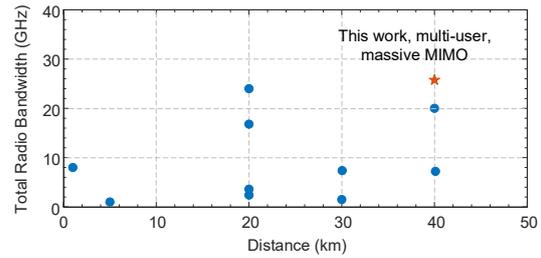

Fig. 1. Summary of reported aggregated radio bandwidth transmitted over a single optical wavelength versus distance for high-capacity IFoF transmissions

To preserve the fully centralized RAN architecture, spectral-efficient IF-over-fiber (IFoF) transmission schemes, where a large number of radio channels can be aggregated and transmitted on a single optical wavelength, have been actively considered [3 - 9]. For example, in [8, 9], 14 radio channels with a total bandwidth of 16.8 GHz (14×1.2 GHz) and 20 channels with a total bandwidth of 20 GHz (20×1 GHz) were aggregated and transmitted over 20 km and 40 km, respectively (summarized in Fig. 1). Even though these demonstrations are highly encouraging [3-9], they are still far away from

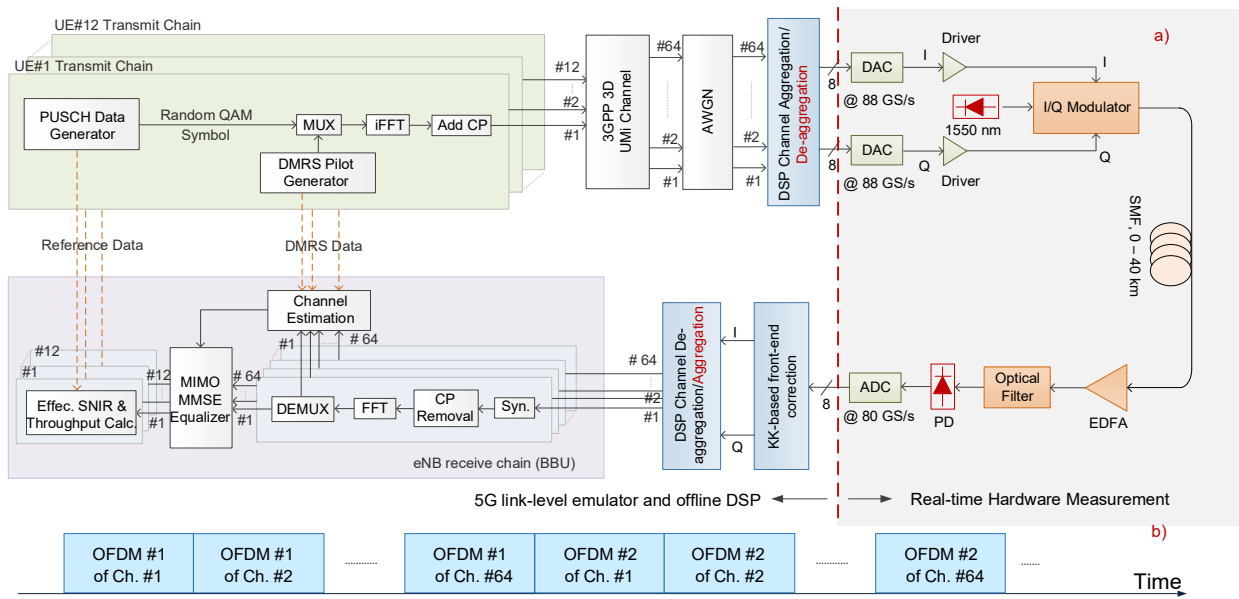

Fig. 2 (a) IFoF transmission setup with 3GPP 3D UMi channel and massive MIMO; (b) Channel aggregation based on TDMA

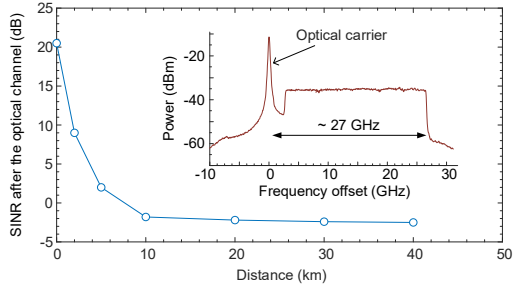

Fig. 3. SINR after optical channel transmission versus link distance; inset shows the optical signal spectrum

practical relevance because (i) they either consider only the fiber-optical channel or they use the wireless channels in the lab with relatively short distances [5, 7], which usually do not represent the noise and fading properties of typical outdoor wireless channels; (ii) the multi-user scenario has been rarely considered.

In this work, we demonstrate an IFoF scheme for 5G massive MIMO RAN with 3GPP 3D Urban Micro (UMi) channel and fully centralized baseband processing. The proposed FH solution allows for the optical channel impairments to be compensated simultaneously with the radio channel response through massive MIMO processing within the baseband receiver. In this case, chromatic dispersion (CD) compensation is not required. Specifically, we consider a 5G RAN with massive MIMO, where 64×400 MHz radio channels are aggregated and transmitted over up to 40 km, achieving a total cell capacity up to 19.4 Gb/s in a multi-user scenario.

**Experimental setup**

The experimental transmission setup is depicted in Fig. 2. For the uplink of a frequency division duplex system, the multi-user 5G-compliant signals were generated through offline digital signal processing (DSP), modelling the uplink scenario of a single cell in a 3GPP 3D Urban Micro deployment [10] with 200 m inter-site distance and 3.5 GHz carrier frequency. We focus here on the up-link scenario since it places a more stringent requirement on the FH transmission as the input signal to FH link is heavily distorted by the wireless propagation channel. The physical uplink shared channel (PUSCH) and DMRS (demodulation reference symbols) signals were generated by a 4096-point FFT, with 30 kHz subcarrier which yields a transmission bandwidth per 5G New Radio (NR) carrier of 100 MHz. We further aggregated four 5G NR carriers, yielding a total transmission bandwidth of 400 MHz. For the DMRS transmission, double symbol mapping type B (i.e., front-loaded) with configuration type 2 was used which allowed the transmission of up to 12 orthogonal layers [10]. For the end-to-end system level performance assessment, we varied the number of users within the cell and the received SNR of each user at the antenna array. The positions of the users were chosen randomly within the sector (assuming a hexagonal deployment), while a closed loop power control ensured the target received SNR at the RRU by controlling the Tx power of the user equipment (UE), assuming no Tx power limitations. The generated time domain baseband signals were then passed through an 3GPP 3D UMi channel model [10] which accurately emulates impulse responses of a typical 5G NR wireless propagation channel (including both azimuth- and elevation angles). The 64 output signals from the 3GPP 3D UMi channel were then sampled at a sampling frequency of 491.52 MHz and fed into the proposed IFoF fronthaul transmission.

These signals were aggregated using a bandwidth efficient time-division multiple access (TDMA) approach as shown in Fig. 2(b) and detailed in [9]. The resulted signal from the aggregation of 64 abovementioned radio channels had a sampling rate of 31.45728 GS/s with a bandwidth of 25.6 GHz. This signal was transmitted over the FH link using a sampling frequency of 32 GS/s. For real-time FH transmission, this signal was digitally resampled to 88 GS/s before being up-converted to an IF of

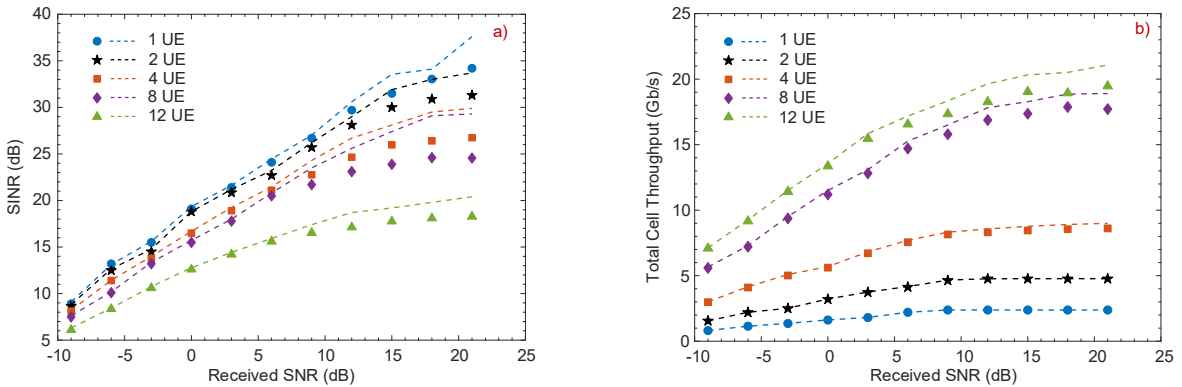

Fig. 4. (a) Post-MIMO SINR and (b) total cell throughput versus received SNR. Dashed lines show the ideal FH case (w/o optical fiber), while markers represent 40 km of SMF FH. The number of UE2 in the cell were 1, 2, 4, 8 and 12.

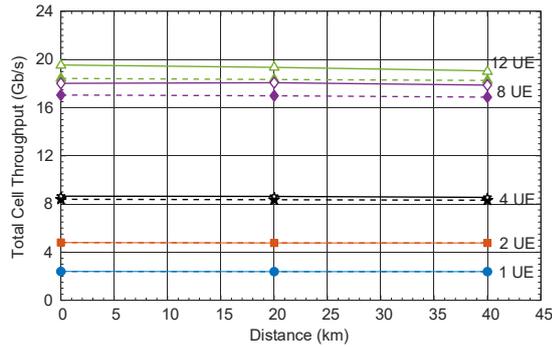

Fig. 5. Total cell throughput versus distance for SNR = 12dB (closed symbols) and SNR = 18 dB (open symbols)

14 GHz to form a complex single-side-band (SSB) signal. Next, the SSB signal was up-converted to the optical domain using two digital-to-analog converters (DACs), two RF drivers, an external cavity laser (ECL) at 1550 nm and an I/Q modulator [9]. The I/Q modulator was off-biased to generate an optical SSB signal as shown in the inset to Fig. 3. The optical bandwidth of this SSB signal was around 27 GHz. The carrier to signal power ratio (CSPR) was fixed at ~ 11 dB. After optical modulation, the optical SSB signal was launched into various spans of standard single mode fiber (SMF) with lengths varying up to 40 km. At the Rx, the signal was amplified by an EDFA, filtered using an optical filter, and detected by a single-ended photodiode (PD). The received signal was then sampled at 80 GS/s by an 8-bit analog-to-digital converter (ADC) and stored offline for further processing.

Before DSP-based channel de-aggregation, we corrected only for the front-end of the optical receiver (PD) by performing the Kramers-Kronig algorithm [11], which removes the signal-signal beat interference resulting from the nonlinearity of the PD. The impact of fiber CD was not compensated. As a result, the signal-to-interference & noise-ratio (SINR) after the optical channel drops below -1 dB even after just 10 km (Fig. 3). After DSP-based channel de-aggregation, 64 separated radio signals were fed into the BBU including channel estimation (CE) and massive MIMO signal processing.

The BBU performed an OFDM symbol timing synchronization and demultiplexed the PUSCH and DMRS data. The latter was used for CE (i.e., pilot-reverse modulation and separation per receive antenna), whose output was used to compute a MIMO-based minimum mean squared error (MMSE) equalizer which mitigates the channel impact and separates the user signals. For each user, the SINR was evaluated per physical resource block (i.e., 12 consecutive subcarriers), which was then averaged over the entire bandwidth of each 400-MHz carrier using the Mutual Information Effective SINR mapping (MIESM, [12]). The effective SINR was then mapped to the effective number of bits per modulation symbol (assuming 256 QAM and LDPC coding [13]), which allowed the computation of a (peak-) data rate per user as well as an aggregate data rate for the cell. We use these performance metrics to quantify the impact of the optical FH system on the performance of the 5G uplink.

**Experimental results and discussions**

The overall system performance is depicted in Fig. 4 for the cases of 1, 2, 4, 8 and 12 UEs over 40 km in comparison with the ideal FH case (w/o optical fiber channel). The received SNR was varied from -9 dB to 21 dB, noting that in practice a received SNR of around 20 dB rarely exist due to Tx power limitation of the UE. At a received SNR < 10 dB, the optical channel at 40 km shows negligible impact on the post-MIMO SINR, for all considered numbers of UEs. When the received SNR is increased beyond 10 dB, the SINR penalty due to the FH channel increases up to a few dB (2 dB-4 dB), for all the numbers of considered UEs in the cell. However, the impact of the FH channel on the total cell throughput behaves differently depending on the number of UEs in the cell. With 1 or 2 UEs, the cell throughput saturates at ~ 2.4 Gb/s and ~ 4.8 Gb/s and almost no impact (<1% of throughput degradation) of the FH link can be observed. When the number of UEs is increased beyond 2, the impact of the FH channel can be noted at high received SNR values. At the maximum considered received SNR value (21 dB) for the case of 4 UEs and 8 UEs, the cell's throughput degradation due to 40 km of FH channel are below 4%, 6.5% and 8%, respectively. However, in the multi-user case, such a high received SNR value rarely happens and thus, does not pose a significant impact on the overall 5G RAN performance. Including 40 km of FH, the maximum cell throughputs are ~ 8.8 Gb/s, 17.7 Gb/s and 19.4 Gb/s for 4, 8 and 12 UEs, respectively. Figure 5 shows that the cell throughput is constant for FH link distances up to 40 km, even though the CD-induced distortion increases rapidly with distance (Fig. 3). We therefore conclude that the fiber CD has been effectively estimated and compensated through the wireless MIMO processing at the BBU. This clearly shows the effectiveness and innovative aspect of the proposed FH solution.

**Conclusion**

We have demonstrated an efficient, joint wireless-optical FH solution for 5G RAN with 3GPP 3D UMi channel and massive MIMO processing. The total aggregated radio bandwidth of 25.6 GHz has been successfully transmitted over 40 km, showing an increase of ~ 28% compared to the previously demonstrated record [8] while no CD compensation is required.